\documentclass[twocolumn,secnumarabic,amssymb,nobibnotes,aps,prd,showpacs,showkeys]{revtex4-1}

\usepackage{graphicx}
\usepackage{amsmath}
\newcommand{\beq}{\begin{equation}}
\newcommand{\eeq}{\end{equation}}
\newcommand{\eml}{\end{mathletters}}

\newcommand{\be}{\begin{equation}}
\newcommand{\ee}{\end{equation}}
\newcommand{\bea}{\begin{eqnarray}}
\newcommand{\eea}{\end{eqnarray}}
\newcommand{\nn}{\nonumber\\}
\newcommand{\oh}{\frac{1}{2}}

\begin{document}
\title{Two-proton emission systematics}

\author{D.S. Delion $^{1,2,3,4}$ and S.A. Ghinescu $^{1,2}$}
\affiliation{
$^1$ "Horia Hulubei" National Institute of Physics and Nuclear Engineering, \\
30 Reactorului, POB MG-6, RO-077125, Bucharest-M\u agurele, Rom\^ania \\
$^2$ Department of Physics, University of Bucharest,
405 Atomi\c stilor, POB MG-11, RO-077125, Bucharest-M\u agurele, Rom\^ania\\
$^3$ Academy of Romanian Scientists, 3 Ilfov RO-050044,
Bucharest, Rom\^ania \\
$^4$ Bioterra University, 81 G\^arlei RO-013724, Bucharest, Rom\^ania}
\date{\today}

\begin{abstract}
The simultaneous emission of two protons is an exotic and complex three-body process.
It is very important for experimental groups investigating the nuclear stability on the 
proton drip line to have a simple rule predicting the two-proton decay widths with a 
reasonable accuracy for transitions between ground as well as excited states in terms 
of relevant physical variables. 
In spite of its complexity, we show that the two-proton emission process obeys similar rules 
as for binary emission processes like proton, alpha and heavy cluster decays.
It turns out that the logarithm of the decay width, corrected by the centrifugal barrier,
linearly depends upon the Coulomb parameter within one order of magnitude. 
On the other hand, the universal linear dependence with a negative slope between the logarithm 
of the reduced width and the fragmentation potential, valid for any kind of binary decay process, 
is also fulfilled for the two-proton emission with a relative good accuracy.
As a consequence of pairing correlations the two protons are simultaneously 
emitted from a singlet paired state. We evidence that indeed one obtains a linear dependence 
between the logarithm of the reduced width and pairing gap within a factor of two, 
giving a good predictive power to this law. It turns out that the diproton and alpha-cluster 
formation probabilities have similar patterns versus the pairing gap, while in the one-proton
case one has a quasi-constant behavior.
\end{abstract}



\maketitle

The proton drip line is mainly investigated by proton and two--proton emission processes 
\cite{Woo97,Son02,Del06a,Pfu12}. The two-proton emission is a very exotic mode that is 
energetically possible in some nuclei. In the earlier sixties Goldansky proposed
two extreme mechanisms in which the particles are emitted, either simultaneouly or 
sequentially \cite{Gol60}. The first systematic theoretical analysis of the processes involving 
the inherent three-body problem was performed in Ref. \cite{Swa65}. 
The theoretical description of two-proton emission was performed by using
few-body formalism in terms of hyperspherical coordinates \cite{Gri01,Gri03,Gri10}, 
as well as R-matrix approaches \cite{Kry95,Bro03}. 
The Feshbach reaction theory and the continuum shell model were also applied \cite{Rot05,Rot06}. 
The one-proton decay systematics reveals simple two-body features depending on the Coulomb and 
centrifugal parameters \cite{Del06}. The systematic analysis indicates that the two-proton emission 
has a three-body character, between the diproton and pure sequential decay \cite{Del09}. 
On the other hand, it is important to point out that the pairing interaction induces a clustering 
of the two protons. This is a fundamental property in $\alpha$-emission, explaining the clustering 
of the four nucleons \cite{Jan83}. In the last years several investigations were perfomed 
in order to describe half-lives of the two-proton emission process by using
effective liquid drop model \cite{Gon17}, 
Gamow coupled channel approach \cite{Wan18},
Gamow model with variable energy \cite{Tav18}, 
semiempirical four-parameter \cite{Str19} 
or two-parameter relation \cite{Liu21}
and the Gamow approach with square nuclear plus Coulomb potential \cite{Liu21a}.

In this paper we will show that the systematics of the two-proton emission has a similar
universal feature compared to the usual binary decays, namely that the logarithm of the reduced width
linearly decreases upon the increase of the fragmentation potential, defined as the difference
between the top of the Coulomb barrier and Q-value. On the other hand, as a consequence
of the pairing corrections between emitted protons, the same quantity is directly proportional
to the pairing gap.

Let us consider the two-proton emission process
\bea
P\rightarrow D+2p~.
\eea
Experimental data evidenced the quasi-simultaneous detection of the emitted protons 
with equal energies. This allowed us to propose in Ref. \cite{Del13} a simplified approach, where 
we have shown that the distribution of emitted protons is centered around the configuration with 
almost equal distances $r_1\sim r_2$, corresponding to polar a angle $\phi\sim\pi/4$ defined by usual relations
\bea
r_1&=&r~cos\phi~,~~~
r_2=r~sin\phi~.
\eea
This is a consequence of the initial condition given by the two-proton pairing wave function 
on the nuclear surface, centered around $\phi=\pi/4$. It is well knon that the deuteron 
has a bound state around $E\sim$ -1 MeV.
The diproton system beyond the nuclear surface becomes unbound, but a simple estimate shows 
that the deuteron bound state is pushed up to a resonant state inside the inter-proton
nuclear plus centrifugal plus Coulomb potential
\bea
\label{poten}
v(r)&=&-v_0\exp\left(-\frac{r^2}{b^2}\right)+\frac{\hbar^2l(l+1)}{2\mu_pr^2}+\frac{e^2}{r}
\nn
v_0&=&35~{\rm MeV}~,~~~b=2~{fm}~,
\eea 
in terms of the relative distance $r=r_1-r_2$, angular momentum $l$ and reduced proton mass $\mu_p=m_p/2$,
at a very small energy $E_{res}\sim$ 0.4 MeV. Therefore the diproton "cluster" is weakly bound
but its center of mass (cm) radius $R=\oh(r_1+r_2)$ moves in the Coulomb field of the daughter nucleus 
as a real diproton particle.
Thus, we suppose for the decay width a similar to the binary case expression, proportional to the scattering amplitude
squared $N^2_L$ in some channel characterized by the angular momentum L \cite{Del10}. It can be rewritten
\bea
\label{Gamma}
\Gamma_L=\hbar v N_L^2=\hbar v \frac{\gamma^2_L(R)}{G_L^2(\chi,\rho)}
\equiv \gamma^2_L(R) P_L(\chi,\rho)~,
\eea
in terms of the reduced width $\gamma^2_L(R)$ and Coulomb penetrability $P_L(\chi,\rho)$.
Let us mention that these quantities differs by a constant factor
with respect to the standard definitions in Ref. \cite{Lan58}.
The penetrability is defined by the irregular Coulomb wave function which has 
the following semiclassical representation
\bea
G_L(\chi,\rho)&=&G_0(\chi,\rho)C_L(\chi,\rho)~.
\eea
Let us mention that the monopole and centrifugal terms are respectively given
\bea
G_0(\chi,\rho)&=&\left(\cot\alpha\right)^{1/2}\exp\left[\chi\left(\alpha-\sin\alpha\cos\alpha\right)\right]
\nn
C_L(\chi,\rho)&=&\exp\left[\frac{L(L+1)}{\chi}\tan\alpha\right]~,
\eea
in terms of the dimensionless parameter
\bea
\cos^2\alpha&\equiv&\frac{\rho}{\chi}=\frac{Q}{V_C(R)}=\frac{QR}{4Z_De^2}~,
\eea
depending upon the Coulomb parameter and reduced radius
\bea
\chi&=&\frac{4Z_De^2}{\hbar v}~,~~~\rho=\kappa R
\nn
v&=&\sqrt{\frac{2Q}{\mu}}~,~~~\hbar\kappa=\mu v~.
\eea
The reduced width $\gamma^2_L$ is also called diproton preformation probability.
We will estimate it on the nuclear surface at the "geometrical touching configuration"
\bea
R=1.2(A_D^{1/3}+A_{2p}^{1/3})~.
\eea
Let us stress that this quantity also includes the dissociation probability of the diproton system. 
In order to prove the validity of this "binary representation" we investigated the available experimental data,
given in Table 1.

\begin{widetext}
\begin{center}
{\it TABLE 1. Parameters of two-proton emission}
\begin{tabular}{|c|c|c|c|c|c|c|c|c|c|c|c|c|c|c|c|}
\hline
no. & $Z_P$ & $N_P$ & $A_P$ & L & $\beta$ & Q &  $V_{frag}$ & $\chi$ & $\rho$ & $\cos^2\alpha$ & $\log_{10}\gamma^2_{exp}$ & $\log_{10}\Gamma_{exp}$ & $\log_{10}\Gamma_{1}$ & $\log_{10}\Gamma_{2}$ & Ref. \cr
 & & & & & & (MeV) & (MeV) & & & & & (MeV) & (MeV) & (MeV) & \cr
\hline
 1 &    4 &    2 &    6 &    0 &    0.000 &    1.370 &    0.316 &    1.248 &    1.014 &    0.813 &   -1.625 &   -1.041 &   -0.562 &   -1.230 & \cite{Wha66} \cr
 2 &    6 &    2 &    8 &    0 &    0.000 &    2.110 &    1.010 &    2.133 &    1.443 &    0.676 &   -1.500 &   -0.881 &   -1.211 &   -1.322 & \cite{Cha11}  \cr
 3 &    7 &    3 &   10 &    1 &    0.250 &    1.300 &    2.381 &    3.508 &    1.239 &    0.353 &   -2.686 &   -3.701 &   -3.254 &   -3.207 & \cite{Gon17} \cr
 4 &    8 &    4 &   12 &    0 &    0.000 &    1.640 &    2.577 &    3.825 &    1.488 &    0.389 &   -0.933 &   -1.241 &   -2.825 &   -2.536 & \cite{Jag12} \cr
 5 &   12 &    7 &   19 &    0 &   -0.240 &    0.750 &    5.514 &    9.769 &    1.170 &    0.120 &   -3.749 &  -10.121 &   -9.439 &   -9.137 & \cite{Muk09} \cr
 6 &   17 &   11 &   28 &    2 &    0.300 &    1.970 &    6.556 &    9.211 &    2.128 &    0.231 &   -3.409 &   -8.391 &   -8.395 &   -7.937 & \cite{Gon17} \cr
 7 &   19 &   13 &   32 &    2 &   -0.120 &    2.080 &    7.262 &   10.208 &    2.273 &    0.223 &   -3.508 &   -9.091 &   -9.097 &   -8.667 & \cite{Gon17} \cr
 8 &   26 &   19 &   45 &    0 &    0.000 &    1.154 &   10.938 &   19.533 &    1.864 &    0.095 &   -3.942 &  -18.941 &  -18.738 &  -18.755 & \cite{Mie07} \cr
 9 &   28 &   20 &   48 &    0 &    0.000 &    1.350 &   11.535 &   19.593 &    2.053 &    0.105 &   -4.699 &  -19.261 &  -18.338 &  -18.426 & \cite{Dos05} \cr
10 &   29 &   23 &   52 &    4 &    0.190 &    0.770 &   12.337 &   26.986 &    1.585 &    0.059 &   -3.918 &  -30.701 &  -30.604 &  -30.794 & \cite{Gon17} \cr
11 &   30 &   24 &   54 &    0 &    0.270 &    1.480 &   11.980 &   20.201 &    2.221 &    0.110 &   -4.125 &  -18.911 &  -18.627 &  -18.732 & \cite{Bla05} \cr
12 &   31 &   26 &   57 &    2 &    0.250 &    2.050 &   11.697 &   17.795 &    2.654 &    0.149 &   -4.047 &  -16.041 &  -15.864 &  -15.889 & \cite{Gon17} \cr
13 &   36 &   31 &   67 &    0 &   -0.270 &    1.690 &   13.763 &   23.041 &    2.520 &    0.109 &   -2.581 &  -19.641 &  -21.011 &  -21.332 & \cite{Goi16} \cr
\hline
\end{tabular}
\end{center}
\end{widetext}

Here we give the charge $Z_P$, neutron $N_P$ and mass number $A_P$ of the parent nucleus, angular momentum
of the emitter diproton $L$, parent quadrupole deformation $\beta$, Q-value, fragmentation potential
\bea
V_{frag}=V_C(R)-Q~, 
\eea
Coulomb parameter $\chi$, reduced radius $\rho=\kappa R$, $\cos^2\alpha$, experimental reduced width
\bea
\label{red}
\gamma^2_{exp}=\frac{\Gamma_0}{P_L}~,
\eea
where the monopole width $\Gamma_0$ is defined below by Eq. (\ref{Gamma0}),
and the logarithm of the experimental decay width $\Gamma_{exp}=\hbar\ln 2/T_{exp}$.
In the last two columns are given two versions for computed decay widths, as described below.

\begin{figure}
\begin{center}
\includegraphics[width=9cm]{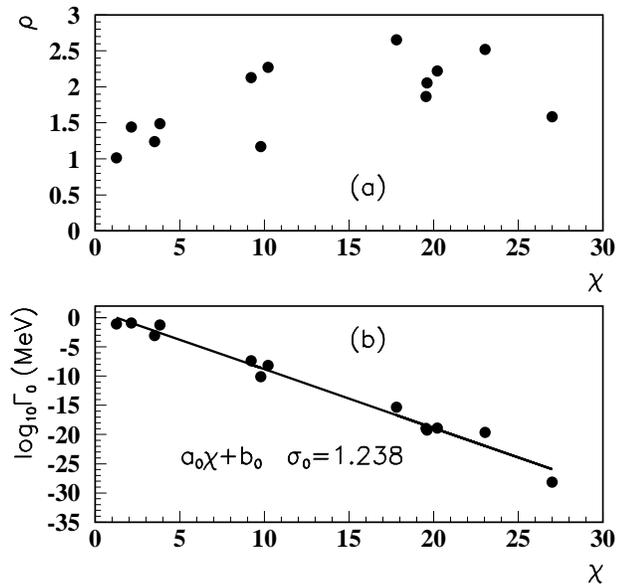} 
\vspace{-5mm}
\caption{
(a) Reduced radius versus Coulomb parameter. \\
(b) Logarithm of the monopole decay width versus the Coulomb parameter.
The fit parameters are given in the first line of the Table 2.
}
\label{fig1}
\end{center} 
\end{figure}

In order to investigate the relation between the two parameters of the Coulomb function
we plot in Fig. \ref{fig1} (a) the dependence between the reduced radius and Coulomb parameter.
Notice their linear correlation except two lower points, corresponding to the cases 4 and 13 in Table 1.
Then we analyzed to what extent the Geiger-Nuttall law, expressing the linear relation between 
the logarithm of the decay with and Coulomb parameter, is fulfilled.
To this purpose we extracted the influence of the centrifugal barrier by defining
the monopole decay width \cite{Del06}
\bea
\label{Gamma0}
\Gamma_0=\Gamma_{exp}C^2_L(\chi,\rho)~.
\eea
In Fig. \ref{fig1} (b) we notice such a linear dependence 
\bea
\log_{10}\Gamma_0\sim a_0\chi+b_0~,
\eea
on a wide interval of almost 30 orders of magnitude. 

In the first line of Table 2 we give the parameters of the fit line with an overall 
root mean square (rms) error $\sigma_0=1.238$. Notice a slightly smaller rms 
error $\sigma_0=1.010$ by excluding the two above mentioned cases 4 and 13. 

\begin{center}
{\it TABLE 2. Parameters of the fit lines}
\begin{tabular}{|c|c|c|c|c|}
\hline
 ~~~$k$~~~ & ~~~$a_k$~~~ & ~~~$b_k$~~~ & ~~~$\sigma_k$~~~ & ~~~$\sigma'_k$~~~ \cr
\hline
 0   &-1.009 & 1.272 & 1.238 & 1.010 \cr
 1   &-0.183 &-1.757 & 0.742 & 0.396 \cr
 2   & 0.622 &-4.876 & 0.702 & 0.333 \cr
\hline
\end{tabular}
\end{center}

\begin{figure}
\begin{center}
\includegraphics[width=9cm]{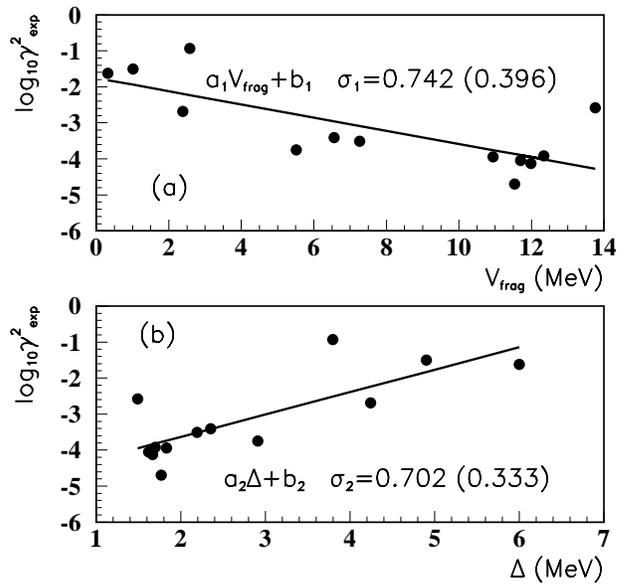} 
\vspace{-5mm}
\caption{
Logarithm of the experimental reduced width at the geometrical touching radius $R=1.2(A_D^{1/3}+2^{1/3})$ 
versus the fragmentation potential (a) and pairing gap (b).
The fit parameters are given in the second line of the Table 2.
The rms error in paratheses corresponds to the analysis without considering cases 4 and 13.
}
\label{fig2}
\end{center} 
\end{figure}

Let us stress that this is a rather large error, coresponding to more than one order of magnitude. 
Therefore this law has an approximate character and a poor predictive power.
This is due to the variation of the reduced width $\gamma^2_L$ along the analyzed emitters.

In order to further analyze this feature let us mention that in all binary emission
processes we evidenced in  Refs. \cite{Del09,Del20} an analytical universal law for reduced widths
\bea
\label{gamma2}
\log_{10}\gamma^2=-\frac{\pi\log_{10}e}{\hbar\omega_1}V_{frag}+\log_{10}s~,
\eea
in terms of the harmonic oscillator frequency of the internal nuclear interaction
approximated by an inverted parabola $\hbar\omega_1$ and the spectroscopic factor $s$.
We plot in Fig. \ref{fig2} (a) the logarithm of the experimental reduced width versus the fragmentation potential. 
Notice that indeed this dependence can satisfactorily be fitted by a straight line
\bea
\log_{10}\gamma^2_{exp}\sim a_1V_{frag}+b_1~,
\eea
with a negative slope $a_1<0$.
The overall rms error in the second line of Table 1 corresponds to a factor of 5
and to a factor of 2.5 if one excludes the cases 4 and 13, corresponding to the upper two points.
The fit parameters lead to the following values in Eq. (\ref{gamma2}) $\hbar\omega_1$=7.456 MeV, $s$=0.017.
Notice that we obtained the same order of magnitude, namely $\hbar\omega_1$=11.389 MeV ($A<145$), 12.580 MeV ($A>145$) 
in the case of the Woods-Saxon potential used to describe one-proton emission and 
$\hbar\omega_1$=9.080 MeV for double-folding potential describing alpha-decay \cite{Dum22}.

\begin{figure}
\begin{center}
\includegraphics[width=9cm]{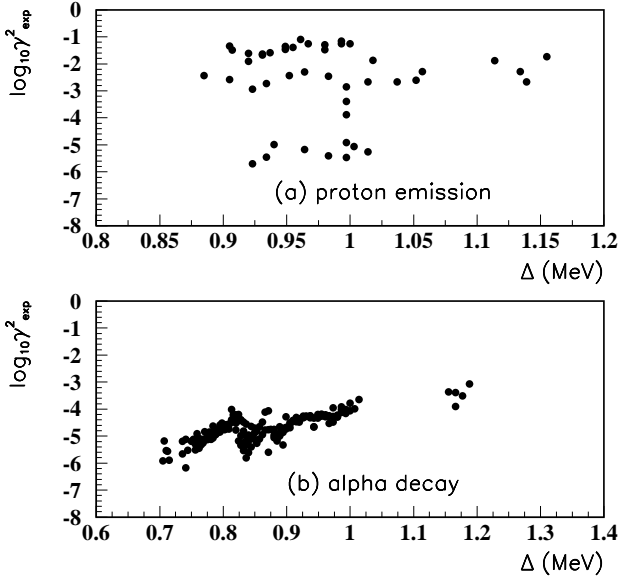} 
\vspace{-5mm}
\caption{
Logarithm of the experimental reduced width versus the pairing gap for proton emission
(a) and alpha decay from even-even emitters (b).
}
\label{fig3}
\end{center} 
\end{figure}

\begin{figure}
\begin{center}
\includegraphics[width=9cm]{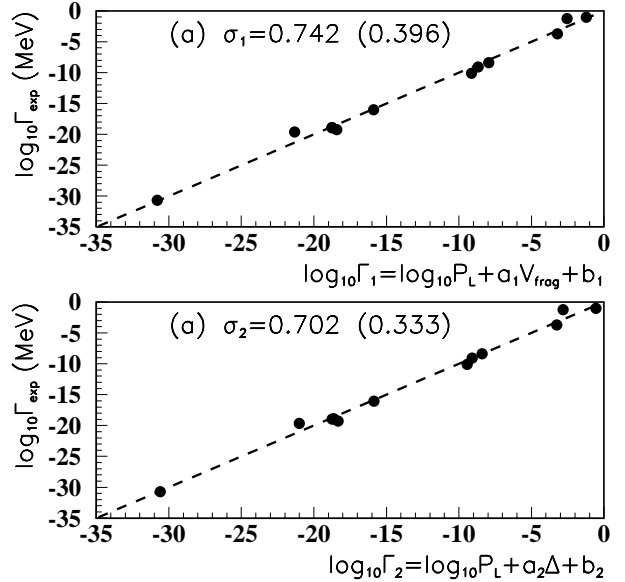} 
\vspace{-5mm}
\caption{
Logarithm of the experimental decay width versus 
$\log_{10}\Gamma_1=\log_{10}P_L+a_1V_{frag}+b_1$ (a) and $\log_{10}\Gamma_2=\log_{10}P_L+a_2\Delta+b_2$ (b).
The fit parameters are given in the third line of Table 2.
The rms error in paratheses corresponds to the analysis without considering cases 4 and 13.
}
\label{fig4}
\end{center} 
\end{figure}

Let us mention here that in Ref. \cite{Del13} we estimated the two-proton formation probability
on the nuclear surface within the pairing BCS approach for $^{45}$Fe (coresponding to no. 8 in Table 1)
as being $\gamma^2_{BCS}\sim 10^{-2}$. From Fig. \ref{fig2} (a) we notice that for $V_{frag}=$ 10.938 MeV
one obtains $\gamma^2_{exp}\sim 10^{-4}$. The missing two orders of magnitude we can ascribe to the
penetration of the diproton system through the inter-proton potential (\ref{poten}).

In the same reference we evidenced that the two-proton formation probability
quadratically depends upon the pairing gap
\bea
\gamma^2_{BCS}\sim (u~v)^2\sim\Delta^2~,
\eea
where $u$ and $v$ denote the standard BCS amplitudes.
Anyway, the systematic analysis of two-proton emitters in Table 1 evidenced that the exponential dependence
upon the pairing gap provides a significantly smaller rms error. 
Thus, we plotted in Fig. \ref{fig2} (b) the dependence between the logarithm of the experimental reduced width 
and the systematic rule of the pairing gap $\Delta=12 A^{-1/2}$
\bea
\log_{10}\gamma_{exp}\sim a_2\Delta+b_2~.
\eea
Notice the linear correlation with a rms error in the third line of Table 2 corresponding to a factor of three.
One obtains a better agreement within a factor of two if one exclude the two mentioned case 4 
(magic in neutrons) and 13.

Let us mention in this context that the situation is quite different in the case of one-proton emission.
In Fig. \ref{fig3} (a) we plotted the experimental reduced width defined by Eq. (\ref{red})
versus the pairing gap. One sees that the upper and lower regions practically do not depend upon the pairing gap.
This is due to the fact that the proton formation probability has an one-particle character, 
being proportional to the BCS amplitude squared $u^2_F$ at the Fermi level.
In the case of the alpha-decay from even-even emitters the behavior is similar to the two-proton emission,
as seen in the panel (b) of the same figure, where we notice two paralel linear dependencies divided by the
doubly magic nucleus $^{208}$Pb.
Therefore the formation probability of the
diproton and alpha cluster has a common collective pairing nature, in spite of the fact
that the first system is unbound, while the second one is strongly bound.

The influence of the quadrupole deformation can be estimated by using the Fr\"oman method 
as in Ref. \cite{Ghi21}. Thus, the influence of the non-diagonal matrix elements
of the Fr\"oman matrix is rather small, being about 15\% for two-proton emitters at $\beta$=0.3.

Finally we analyzed the experimental decay width by plotting in Fig. \ref{fig4} its logarithm 
as a function of the theoretical width for the panels (a) and (b), respectivelly
\bea 
\log_{10}\Gamma_1&=&\log_{10}P_L+a_1V_{frag}+b_1~,~~~{(a)} 
\nn
\log_{10}\Gamma_2&=&\log_{10}P_L+a_2\Delta+b_2~,~~~{(b)}~. 
\eea
The values of the theoretical predictions are given in the last two columns of Table 1 and
the fit parameters in the last two lines of Table 2.
Notice the very good linear correlation between these quantities along the first bisectrice
plotted by a dashed line.

Concluding, in spite of the fact that the two-proton emission is a three-body process, 
we evidenced the binary character of laws connecting the logarithm of the decay widths in terms
some physical quantities. Thus, we evidenced the linear correlation between the logarithm
of the reduced width and the fragmentation potential with a negative slope, predicted
as an analytical universal rule for binary emission processes like proton emission,
alpha and heavy cluster decays. On the other hand, we also evidenced the role played
by the pairing interaction, given by the linear correlation between the logarithm of 
the reduced width and pairing gap, as predicted by microscopic estimates of the two-proton 
formation probability. The relative small rms error give a powerfull predictive power to this last rule.
The diproton and alpha-cluster formation probabilities have similar paterns versus the pairing gap, 
while in the one-proton emission one has a quasi-constant behavior.

\begin{acknowledgments}
This work was supported by the grant of the Romanian
Ministry Education and Research PN-18090101/2019-2021 and by the grant
01-3-1136-2019/2023 of JINR-Dubna.
\end{acknowledgments}



\begin{thebibliography}{99}
\bibitem{Woo97} P. J. Woods and C. N. Davids, 
Annu. Rev. Nucl. Part. Sci. {\bf 47}, 541 (1997).
\bibitem{Son02} A. A. Sonzogni, 
Nuclear Data Sheets {\bf 95}, 1 (2002).
\bibitem{Del06a} D.S. Delion, R.J. Liotta, R. Wyss,
Phys. Rep. {\bf 424}, 113 (2006).
\bibitem{Pfu12} M. Pf\"utzner, M. Karny, L.V. Grigorenko, and K. Riisager,
Rev. Mod. Phys. {\bf 84}, 567 (2012).
\bibitem{Gol60} V. I. Goldansky, Nucl. Phys. {\bf 19}, 482 (1960).
\bibitem{Swa65} P. Swan, Rev. Mod. Phys. {\bf 37}, 336 (1965).
\bibitem{Gri01} L. V. Grigorenko, R. C. Johnson, I. G. Mukha, I. J. Thompson, and M. V. Zhukov, 
Phys. Rev. C {\bf 64}, 054002 (2001).
\bibitem{Gri03} L. V. Grigorenko and M. V. Zhukov, Phys. Rev. C {\bf 68}, 054005 (2003), and references therein.
\bibitem{Gri10} L. V. Grigorenko, I. A. Egorova, M. V. Zhukov, R. J. Charity,
and K. Miernik, Phys. Rev. C {\bf 82}, 014615 (2010).
\bibitem{Kry95} R. A. Kryger, A. Azhari, M. Hellstrom, J. H. Kelley, T. Kubo, R. Pfaff, {\it et al.}, 
Phys. Rev. Lett. {\bf 74}, 860 (1995).
\bibitem{Bro03} B. A. Brown and F. C. Barker, Phys. Rev. C {\bf 67}, 041304(R) (2003), and references therein.
\bibitem{Rot05} J. Rotureau, J. Okolowicz, and M. Ploszajczak, Phys. Rev. Lett. {\bf 95}, 042503 (2005).
\bibitem{Rot06} J. Rotureau, J. Okolowicz, and M. Ploszajczak, Nucl. Phys. A {\bf 767}, 13 (2006).
\bibitem{Del06} D. S. Delion, R. J. Liotta, and R. Wyss, Phys. Rev. Lett. {\bf 96}, 072501 (2006).
\bibitem{Del09} D. S. Delion, Phys. Rev. C {\bf 80}, 024310 (2009). 
\bibitem{Jan83} F. A. Janouch and R. J. Liotta, Phys. Rev. C 27, 896 (1983).
\bibitem{Gon17} M. Gonalves, N. Teruya, O. Tavares, and S. Duarte, Phys. Lett. B {\bf 774}, 14 (2017).
\bibitem{Wan18} S. M. Wang, W. Nazarewicz, Phys. Rev. Lett. {\bf 120}, 212502 (2018).
\bibitem{Tav18} O. A. P. Tavares, E. L. Medeiros, Eur. Phys. J. A {\bf 54}, 65 (2018).
\bibitem{Str19} I. Sreeja and M. Balasubramaniam, Eur. Phys. J. A {\bf 55}, 33 (2019).
\bibitem{Liu21} H. M. Liu, Y. T. Zou, X Pan et al., Chin. Phys. C {\bf 45}, 024108 (2021).
\bibitem{Liu21a} H. M. Liu, {\it et. al.} Chin. Phys. C {\bf 45}, 043110 (2021).
\bibitem{Del13} D. S. Delion, R. J. Liotta, and R. Wyss, Phys. Rev. C {\bf 87}, 034328 (2013).
\bibitem{Del10} D. S. Delion, Theory of Particle and Cluster Emission (Springer-Verlag, Berlin, New York, 2010).
\bibitem{Lan58} A. M. Lane and R. G. Thomas, Rev. Mod. Phys. {\bf 30}, 257 (1958).
\bibitem{Wha66} W. Whaling, Phys. Rev. {\bf 150}, 836 (1966).
\bibitem{Cha11} R. J. Charity, J.M. Elson, J. Manfredi, R. Shane, L. G. Sobotka, B.A. Brown, {\it et al.}, 
Phys. Rev. C {\bf 84}, 014320 (2011).
\bibitem{Jag12} M.F. Jager, R. J. Charity, J. M. Elson, J. Manfredi, M. H. Mahzoon, L. G. Sobotka, {\it et al.}, 
Phys. Rev. C {\bf 86}, 011304(R) (2012).
\bibitem{Muk09} I. Mukha for the S271 Collaboration, Eur. Phys. J. A {\bf 42}, 421 (2009).
\bibitem{Mie07} K. Miernik, W. Dominik, Z. Janas, M. Pfutzner, L. Grigorenko, C. R. Bingham, {\it et al.}, 
Phys. Rev. Lett. {\bf 99}, 192501 (2007).
\bibitem{Dos05} C. Dossat, A. Bey, B. Blank, G. Canchel, A. Fleury, J. Giovinazzo, {\it et al.}, 
Phys. Rev. C {\bf 72}, 054315 (2005).
\bibitem{Bla05} B. Blank, A. Bey, G. Canchel, C. Dossat, A. Fleury, J. Giovinazzo, {\it et al.}, 
Phys. Rev. Lett. {\bf 94}, 232501 (2005).
\bibitem{Goi16} T. Goigoux {\it et al.}, Phys. Rev. Lett. {\bf 117}, 162501 (2016).
\bibitem{Del20} D. S. Delion and A. Dumitrescu, Phys. Rev. C {\bf 102}, 014327 (2020).
\bibitem{Dum22} A. Dumitrescu and D.S. Delion, At. Data Nucl. Data Tables (2022) (in press).
\bibitem{Ghi21} S. A. Ghinescu and D. S. Delion, J. Phys. G {\bf 48}, 105108 (2021).
\end{thebibliography}
\end{document}